\documentclass[reprint,aps,pra,twocolumn, superscriptaddress,floatfix]{revtex4-2}
\usepackage{amsmath}
\usepackage{graphicx}
\usepackage[colorlinks=true,linkcolor=blue]{hyperref}
\usepackage{appendix}
\usepackage{array}
\usepackage{multirow}

\begin{document}

\title{Reply to ``Comment on `Generalized James' effective Hamiltonian method'"}
\author{Wenjun Shao}
\affiliation{Department of Physics, School of Science, Westlake University, Hangzhou 310030,  China}
\affiliation{Institute of Natural Sciences, Westlake Institute for Advanced Study, Hangzhou 310024, China}
\author{Chunfeng Wu}
\email{chunfeng_wu@sutd.edu.sg}
\affiliation{Science, Mathematics and Technology, Singapore University of Technology and Design, 8 Somapah Road, Singapore 487372}
\author{Xun-Li Feng}
\email{xunli_feng@sutd.edu.sg}
\affiliation{Department of Physics, Shanghai Normal University, Shanghai 200234, China}

\begin{abstract}
In the preceding Comment \cite{comment} it was claimed that the third-order Hamiltonian obtained in our original paper \cite{Shao} is not Hermitian for general situations when considering time-dependence and the way of deriving the effective third-order expansion is not very rigorous. To reply the comment we should emphasize the following three points: first of all, the third-order Hamiltonian given in our paper is exactly Hermitian under the conditions mentioned there. Secondly, the iterative method adopted in our paper to derive the generalized effective Hamiltonian is equivalent to the Dyson series, and its correctness can thus be guaranteed. Thirdly, although the truncated effective Hamiltonian is indeed non-Hermitian under the time-dependent situation as presented in the Comment, it corresponds exactly to the non-unitary truncated Dyson series. Considering the truncated Dyson series has been extensively utilized in the time-dependent perturbation theory, in our opinion, the non-Hermitian truncated effective Hamiltonian can still be treated as an approximation of the effective Hamiltonian.
\end{abstract}

\maketitle

 In 2000, James proposed an effective Hamiltonian method in
the appendix of his paper \cite{James-2000}, which provides us an efficient tool to solve many interesting questions related to light-matter interactions with large detuning. Such a method was extensively cited and was later referred to as James' effective Hamiltonian method. This method is usually applicable to deal with the following Hamiltonian in the interaction picture
\begin{equation}
\hat{H}_{\text{I}}(t)=\sum_{m}\left( \hat{h}_{m}e^{i\omega _{m}t}+\hat{h}%
_{m}^{\dag }e^{-i\omega _{m}t}\right) ,  \label{H_I}
\end{equation}%
where the operators $\hat{h}_{m}$ are generally independent of time. By formally solving the Schr\"{o}dinger equation governed by Eq. \eqref{H_I} and resorting to the iterative method, one can arrive at an effective Hamiltonian under the Markovian approximation, 
\begin{equation}
\hat{H}_{\text{eff}}^{(2)}=\frac{1}{i\hbar }\hat{H}_{\text{I}}(t)\int_{0}^{t}%
\hat{H}_{\text{I}}(t^{\prime })dt^{\prime }.  \label{Heff2}
\end{equation}

If all the frequencies $\omega _{m}$ are distinct, substituting Eq. \eqref{H_I} into Eq. \eqref{Heff2} yields the following effective Hamiltonian under the rotating wave approximation in which all the terms containing $\exp [\pm i(\omega_{m}-\omega _{n})]$ are neglected because $\omega _{m}-\omega _{n}$ is still large enough to meet the rotating wave approximation 
\begin{equation}
\hat{H}_{\text{I}}(t)=\sum_{m}\frac{1}{\hbar \omega _{m}}[\hat{h}_{m},\hat{h}%
_{m}^{\dag }].  \label{H3}
\end{equation}

If the condition that all the frequencies $\omega _{m}$ are distinct is not satisfied, which means two or more frequencies, say $\omega _{m}$ and$\omega _{n}$, are close in quantity, which means $\left\vert \omega_{m}-\omega _{n}\right\vert \ll \omega _{m},\omega _{n},$ the terms containing $\exp [\pm i(\omega _{m}-\omega _{n})]$ are not negligible because $\omega _{m}-\omega _{n}$ is small and the rotating wave approximation cannot be satisfied. In this case the resultant effective Hamiltonian might be non-Hermitian. To overcome such problems James and his colleagues improved the original James' effective Hamiltonian method by using a time-averaged filter operation \cite{James-2007,James-2010}.

In our paper \cite{Shao} we gave a generalized version of James' effective Hamiltonian method \cite{James-2000} by using the iterative method and obtained an effective Hamiltonian of the form 
\begin{equation}
\hat{H}_{\text{eff}}(t)=\hat{H}_{\text{eff}}^{(2)}(t)+\hat{H}_{\text{eff}%
}^{(3)}(t)+\cdots +\hat{H}_{\text{eff}}^{(n)}(t)+\cdots ,  \label{Heff}
\end{equation}%
where $\hat{H}_{\text{eff}}^{(2)}(t)$ takes the form of Eq.  \eqref{Heff2} and $\hat{H}_{\text{eff}}^{(3)}(t)\ $takes the following form 
\begin{equation}
\hat{H}_{\text{eff}}^{(3)}=\left( \frac{1}{i\hbar }\right) ^{2}\hat{H}_{\text{I}}(t)\int_{0}^{t}\hat{H}_{\text{I}}(t_{1})\int_{0}^{t_{1}}\hat{H}_{\text{I}}(t_{2})dt_{2}dt_{1}.  \label{Heff3}
\end{equation}%
For more details and the form of $n$th order $\hat{H}_{\text{eff}}^{(n)}(t)$
readers can refer to our paper \cite{Shao}.

As shown in Ref. \cite{Shao}, the effective Hamiltonian expressed in Eq. \eqref{Heff} is equivalent to the Dyson series \cite{Dyson}. So the iterative method adopted to derive the Eq. \eqref{Heff} is no problem and the correctness of our results can be guaranteed. As one knows, the Dyson series is extensively employed in many fields such as quantum mechanics and quantum theory of fields, and the truncated sum of the Dyson series is also a basis for time-dependent perturbation theory. From this point of view, one can reasonably recognize the truncated sum of the front items in Eq. \eqref{Heff} as an approximation of the effective Hamiltonian.

We should point out that, under the condition 
$\left[ \hat{H}_{I}(t_{1}),\hat{H}_{I}(t_{2})\right] \neq 0$ for arbitrary time $t_{1}$ and $t_{2}$,
the truncated sum of the Dyson series is usually non-unitary \cite{ME}.
However, such a non-unitary character does not mean that the Dyson series is useless. On the contrary, it has been successfully utilized in quantum mechanics and quantum theory of fields. Correspondingly, the terms $\hat{H}_{\text{eff}}^{(n)}(t)$ ($n=2,3,\cdots $) and thus their truncated sum in Eq. \eqref{Heff} might be non-Hermitian under the same condition $\left[ \hat{H}_{I}(t_{1}),\hat{H}_{I}(t_{2})\right] \neq 0$. Nonetheless, we think the truncated sum of the front items in Eq. \eqref{Heff} can still be treated as an approximation of the effective Hamiltonian considering the fruitful application of the truncated sum of the Dyson series.

In the Comment the authors showed that the third-order effective Hamiltonian (15) in our paper \cite{Shao} is not Hermitian under the general situations when considering time-dependence. This result is not surprising according to the above discussion. But the way they \cite{comment} reasoned from Eq. (6) to Eq. (8) is misleading because the truncated sum of the Dyson series does not meet the unitary evolution requirement expressed in Eq. (7) any more. In addition, even under the unitary evolution, that is, the Eq. (7) is met, it seems that one cannot deduce the Eq. (8) directly from $H_{eff}^{(3)\dagger}=H_{eff}^{(3)},$ which implies 
\begin{widetext} 
\begin{equation}
\hat{H}_I(t) \int_{0}^{t}dt_1 \int_{0}^{t1}dt_2 \hat{H}_I(t_1) \hat{H}_I(t_2)=  \int_{0}^{t}dt_1 \int_{0}^{t1}dt_2 \hat{H}_I(t_2) \hat{H}_I(t_1) \hat{H}_I(t).  \label{H3+=H3}
\end{equation} 
\end{widetext}
Considering $\hat{H}_{I}(t)$ is in the left of the left side while in the right of the right side of the Eq. \eqref{H3+=H3} and $\left[ \hat{H}_{I}(t),\hat{H}_{I}(t_{1})\right] \neq 0$ and $\left[ \hat{H}_{I}(t),\hat{H}_{I}(t_{2})\right] \neq 0$, one cannot directly obtain $\hat{H}_{I}(t_{1})\hat{H}_{I}(t_{2})=\hat{H}_{I}(t_{2})\hat{H}_{I}(t_{1})$ from Eq. \eqref{H3+=H3}. 
Thus, the result Eq. (8) in the comment \cite{comment} is not correct, neither are the following analyses.

Moreover, we should point out that the third-order effective Hamiltonian (15) in \cite{Shao} is exactly Hermitian under the situation that all of the frequencies $\omega _{m}$ are distinct as stated in our paper \cite{Shao}:\textquotedblleft we mainly focus on the third-order case and limit ourselves to the case that all of the frequencies $\omega _{m}$ are distinct and the algebraic sum of any three frequencies, including two same ones, is equal to zero or distinct from zero", the Hermiticity of the third-order effective Hamiltonian (15) in \cite{Shao} was also proved in the appendix under the conditions mentioned.

The third-order Hamiltonian expressed in Eq. (19) in the comment was derived by using the time-averaged dynamics method developed in Ref. \cite{James-2007}, which can avoid the non-Hermitian effective Hamiltonian, but this does not mean our result is not correct. In fact, our iterative method is equivalent to that of the Dyson series, and the non-Hermitian character of the truncated effective Hamiltonian is just corresponding to the non-unitary character of the truncated Dyson series. The third-order effective Hamiltonian (15) in Ref. \cite{Shao} is exactly Hermitian under the conditions mentioned in our paper, and thus it can be safely used to solve problems which meet these conditions. \\

In addition, as an approximate method, the third-order effective Hamiltonian  corresponding to the third-order perturbation theory can be regarded as Hermitian when the algebraic sum of any three frequencies infinitely \emph{approaches to zero, but not equal to zero}.
Because the coefficients of the term and its Hermitian part in $ \hat{H}_{\text{eff}}^{(3)} $ are approximately equal under such a condition. 
For example, in Sec. IV of the comment \cite{comment} the authors obtain Eq. (25) according to our paper \cite{Shao}, where terms with $ \delta=\Delta_1+\Delta_2-\omega $ approaching to zero in the exponent are kept and the other rapid oscillating terms are all neglected according to the rotating wave approximation. 
After substituting the relation $ \delta=\Delta_1+\Delta_2-\omega $ into Eq. (25), one may find 
\begin{equation}
\zeta_{eff} \approx \zeta'_{eff} \quad \text{and} \quad  \xi_{eff} \approx \xi'_{eff}
\end{equation}
when $ \delta $ is close to zero. Therefore, the effective time-dependent Hamiltonian (25) can be regarded as Hermitian.
Under the condition mentioned in our paper \cite{Shao}, i.e.,  $ \delta=0 $, the both give the same results.

By the way, in the comment \cite{comment} the Eq. (26) is not very exact, the correct form is 
\begin{equation*}
\hat{H}^{(2)}_0= \dfrac{g^2_2}{\Delta_2}\hat{a}^{\dagger}\hat{a}\hat{\sigma}_{ee}-\dfrac{g^2_1}{\Delta_1}\hat{a}^{\dagger}\hat{a}\hat{\sigma}_{gg}-\dfrac{\lambda^2}{\omega} + \left( \dfrac{g^2_1}{\Delta_1}-\dfrac{g^2_2}{\Delta_2}\right) \hat{a}\hat{a}^{\dagger}\hat{\sigma}_{ii},
\end{equation*} 
which lacks the last term that the authors discard.
Moreover, the Eq. (27) is not very exact, which can be further simplified,  and the correct form is 
\begin{equation*}
\hat{H}=\hat{H}^{(2)}_0+g_{eff} \left( \hat{a}^{\dagger} \hat{\sigma}_{ge} + \hat{a}\hat{\sigma}_{eg} \right).
\end{equation*}
Furthermore, it is obvious that by applying a unitary transformation with $ \hat{U}=\text{exp}\left[ -i\hat{H}^{(2)}_0 t \right] $ on the Hamiltonian (28) one cannot directly obtain Eq. (29), which has discarded many terms without any assumptions and approximations.

At last, the authors make some misunderstandings and mistakes about James' work \cite{James-2007, James-2010}.
For example, following the Sec. II D of Ref. \cite{James-2010} due to Ref. \cite{James-2007}, the authors give an addition part III B in their comment \cite{comment}. At the beginning of this part they take $ \overline{\hat{H}_I}=0 $ into Eq. (A6) in Ref. \cite{James-2010} to find Eqs. (20)--(23) in the comment \cite{comment}, which lacks the generality, this just shows a special condition for Hamiltonians with harmonic time dependence. 
% otherwise Eqs. (17) -- (22) in Sec. II D \cite{James-2010} should be further reduced.
Without loss of generality, from Eq. (A6) one may get
\begin{equation}
\mathcal{L}_3[\overline{\rho}] =\left[\hat{H}_{eff}^{(3)},\overline{\rho} \right] + \left\lbrace  \dfrac{\hat{B}-\hat{B}^{\dagger}}{2},\overline{\rho} \right\rbrace  + \mathcal{D}_3[\overline{\rho}],  
\end{equation}
where the effective Hamiltonian 
\begin{equation}
\hat{H}_{eff}^{(3)} = \dfrac{\hat{B}+\hat{B}^{\dagger}}{2}  \label{H3_eff}
\end{equation}
with 
\begin{equation*}
\hat{B}=  \overline{\hat{H}_I \hat{U}_2}- \overline{\hat{H}_I} \ \overline{\hat{U}_2 } - \overline{\hat{H}_I \hat{U}_1} \  \overline{\hat{U}_1} + \overline{\hat{H}_I} \ \overline{\hat{U}_1 } \ \overline{\hat{U}_1},
\end{equation*}
and the decoherence terms
\begin{equation}
\begin{split}
\mathcal{D}_3[\overline{\rho}] = &  \overline{\hat{H}_I \rho \hat{U}^{\dagger}_2} -  \overline{\hat{H}_I} \rho \overline{ \hat{U} ^{\dagger}_2} -  \overline{\hat{U}_2 \rho \hat{H}_I} +  \overline{\hat{U}_2} \rho  \overline{\hat{H}_I} \\
& + \overline{\hat{H}_I \hat{U}_1 \rho \hat{U}^{\dagger}_1} - \overline{\hat{H}_I \hat{U}_1} \rho \overline{\hat{U}^{\dagger}_1}  - \overline{\hat{H}_I \overline{\hat{U}_1 } \rho \hat{U}^{\dagger}_1} \\
& -    \overline{\hat{H}_I} \ \overline{\hat{U}_1 \rho \hat{U}^{\dagger}_1} + 2  \overline{\hat{H}_I} \ \overline{\hat{U}_1 }\rho \overline{\hat{U}^{\dagger}_1} -  \overline{\hat{U}_1 \rho \hat{U}^{\dagger}_1 \hat{H}_I} \\
& + \overline{\hat{U}_1} \rho \overline{\hat{U}^{\dagger}_1 \hat{H}_I} +\overline{\hat{U}_1 \rho \overline{\hat{U}^{\dagger}_1} \hat{H}_I}  + \overline{\hat{U}_1 \rho \hat{U}^{\dagger}_1 } \  \overline{\hat{H}_I} \\
&- 2 \overline{\hat{U}_1} \rho \overline{\hat{U}^{\dagger}_1 } \  \overline{\hat{H}_I} - \overline{ \hat{H}_I \rho \overline{\hat{U}^{\dagger}_1} \hat{U}^{\dagger}_1} + \overline{ \hat{H}_I} \rho \overline{\hat{U}^{\dagger}_1}  \ \overline{\hat{U}^{\dagger}_1}\\
& + \overline{\hat{U}_1 \overline{\hat{U}_1} \rho  \hat{H}_I } - \overline{\hat{U}_1} \ \overline{\hat{U}_1} \rho \overline{\hat{H}_I }.
\end{split}
\end{equation}
are different from those in the comment \cite{comment}. By applying the above general result to a class of harmonic time-dependent Hamiltonians, %i.e., Eq. \eqref{H_I}, 
 one will find that Eq. \eqref{H3_eff} leads to Eq. (18) in the comment.

\newpage

\section{Reports of the Referee and reply}

\subsection{Reports}

Re: AHK1043 

Reply to ``Comment on ``Generalized James' effective Hamiltonian method'" by Wenjun Shao, Chunfeng Wu, and Xun-Li Feng\\

Dear Dr. Feng,\\

The above manuscript has been reviewed by one of our referees. Comments from the report appear below.

These comments suggest that the present manuscript is not suitable for publication in the Physical Review.\\

Yours sincerely,

Dr. *** ***

Managing Editor

Physical Review A\\
--------------------------------------------------------------------------
Report of the Referee -- AHK1043/Shao \\
--------------------------------------------------------------------------

After reading the revised comment paper again, I was unsure whether the authors of the reply have read it. As far as I can see, the authors of the comment paper acknowledge that the original method does produce Hermitian Hamiltonian for time independent cases. This is clear in the abstract ``...here we show that the third-order Hamiltonian obtained in [1] is not Hermitian for general situations when we consider time-dependence." and in the main text ``In [1], it was proved that the Hamiltonian is Hermitian for a particular case where the effective Hamiltonian is independent of time, but not in general.".

I could not find where the comment paper ``ignored the conditions" and
claimed that the result is wrong. I think it is fair to point out that
the time-dependent Hamiltonian is not Hermitian and there's a way to
fix it. This does not mean that the result of the original paper is wrong in its regime of validity.
 
As the current reply does not bring in new physics or arguments (other than those already explained in the comment paper), I cannot recommend
the reply paper for publication in Physical Review A.

\newpage

\subsection{Reply}

Dear Editor,\\

In the referee's report it was emphasized that the authors of the comment paper acknowledge the original method does produce Hermitian Hamiltonian for time independent cases. But the fact is not the case, for example, in the abstract ``...here we show that the third-order Hamiltonian obtained in [1] is not Hermitian for general situations when we consider time-dependence." In the main text of the comment ``In [1], it was proved that the Hamiltonian is Hermitian for a particular case where the effective Hamiltonian is independent of time, but not in general."
In our understanding, these comments indeed negate the conclusions of our paper based on an exaggerated condition of  time-dependent Hamiltonian. So it is absolutely necessary to clarify such a point in a reply.

Moreover, the referee did not think the comment paper ``ignored the conditions" and claimed that the result is wrong and the referee thought “...This does not mean that the result of the original paper is wrong in its regime of validity.” 
Our response is, that being the case, why did the authors write this comment paper based on exaggerated conditions beyond our original paper, instead of a regular paper as suggested in our early response: ``Therefore, we think that this manuscript cannot be accepted for publication in Phys. Rev. A as a comment. Otherwise, the result presented in Sec. III of the manuscript is interesting and can be regarded as an extension of Ref. [2]. So, it might be considered for publication in Phys. Rev. A as a regular article if the following questions are well addressed. ”

In conclusion, as stated in our initial response, we believe that since our original paper was correct in the claimed conditions, the current comment paper loses its legitimacy for publication because the conditions of our original paper was exaggerated, so we still recommend that this work be published as a regular paper. If it is still published in the current version, our reply should be allowed to be published.\\

Sincerely yours,

The authors

\newpage
\section{Reports and replies}

\subsection{Reports}

Dear Dr. Feng,\\

We acknowledge receipt of your letter. What you write about the
Comment is, respectfully, immaterial; you have had a chance to state
your position before the Comment underwent anonymous review, it has
now been reviewed and recommended for publication, and that is that.
You should, instead, focus on the content of your Reply. If you
disagree with the judgment of the referee, then you should explain
your reasons for doing so. This should not be a statement about the
alleged (un)fairness of the process, since publishing a Comment
without a Reply is perfectly in line with our policies; see, e.g., the
final sentence of the appended memo. Instead, you should focus on
explaining in what way the Reply adds something new to the discussion.\\

Yours sincerely,

Dr. *** ***

Managing Editor

Physical Review A

Email: pra@aps.org

https://journals.aps.org/pra/ \bigskip \\
--------------------------------------------------------------------------
\bigskip
      
Dear Dr. Feng,\\

In order to proceed with consideration of this manuscript, we require
a more complete resubmittal letter indicating in detail your responses
to the referee report(s) and the changes made to the manuscript, if
any. We await your response.\\

Yours sincerely,

*** ***

\newpage

\subsection{Reply}      
      
Dear editor,\\

In the new reply we bring in new arguements, which mainly contains two points. 
One is the non-Hermiticity of the third-order effective Hamiltonian. Our generalized effective Hamiltonian derived by the iterative method is equivalent to the Dyson series, whose correctness can thus be guaranteed. Although the truncated effective Hamiltonian is indeed non-Hermitian under the time-dependent situation as presented in the Comment, it corresponds exactly to the non-unitary truncated Dyson series, which has been extensively utilized in the time-dependent perturbation theory, in our opinion, the non-Hermitian truncated effective Hamiltonian can still be treated as an approximation of the effective Hamiltonian. Besides, as an approximate method, the third-order effective Hamiltonian can be regarded as Hermitian after rotating wave approximation, because the effective coefficient/coupling rate and its Hermitian part are approximately equal. 
The other is the deriving of Eq. (8), which cannot be directly obtained from Eq.  (6).  The result can be easily found by taking a commutation relation for different time t1 and t2.
 
We think the added new discussion can can fulfill the requirement and should be published. \\

Yours sincerely,

The authors\\
      
\bigskip

The new manuscript is attached on arXiv:
2312.05732v2, {\color{blue} https://arxiv.org/abs/2312.05732v2  }

\newpage

\section{Reports   and   replies}

\subsection{Reports}

Re: AHK1043 

Reply to ``Comment on ``Generalized James' effective Hamiltonian  method'" by Wenjun Shao, Chunfeng Wu, and Xun-Li Feng\\

Dear Dr. Feng,\\

Your Reply referenced above has been reviewed again by the previous
referee. In view of the fact that the referee is still negative, we
subsequently sent the Reply also to a new referee for an additional
opinion. Comments from both reports are appended below.

We regret that in view of these comments we cannot accept the paper
for publication in the Physical Review. This concludes our review of
your manuscript; no further revisions of the manuscript can be
considered.\\

Yours sincerely,

Dr. *** ***

Managing Editor

Physical Review A

Email: pra@aps.org

https://journals.aps.org/pra/

Follow us on Twitter @PhysRevA

P.S.  We regret the delay in obtaining these reports. \\
--------------------------------------------------------------------------
Second Report of the First Referee -- AHK1043/Shao\\
--------------------------------------------------------------------------

I'm afraid that I still do not see any new physics or arguments. The
main point now seems to be that the method is equivalent to the
truncated Dyson series method, but this is already mentioned in the
comment paper. I still fail to see the merit in publishing the
manuscript, but perhaps a new set of eyes are needed in this case.

Just one comment. I do not quite follow the argument below Eq. (6).
Doesn't the condition hold if the Hamiltonian at two different time
commute?\\ 
--------------------------------------------------------------------------
Report of the Second Referee -- AHK1043/Shao\\
--------------------------------------------------------------------------

Non-unitary evolution due to non-Hermitian effective Hamiltonians is
an issue and this is why the Comment paper tries to fix the problem.
While the authors of the Reply claim that the non-Hermitian nature of
their effective Hamiltonians is ``no problem", they fail to provide
strong reasons to justify their view. Specifically, when time
evolution of quantum states is a topic of interest (for example in
quantum optics), an accurate description of the states must require
unitary evolution. I think that the argument based on truncated Dyson
series in the Reply does not really address the issue. Since the Reply
does not bring in something sufficiently meaningful and worthwhile to
the discussion, I do not recommend its publication.

Finally, I would like to add a remark:

In the Reply the authors are indeed correct that Eq. 8 of the Comment
paper cannot be obtained from the condition  H\_eff\string^3+= H\_eff\string^3, 
although Eq. 8 guarantees H\_eff\string^3+= H\_eff\string^3. I think that this is a
minor problem that does not affect the main part of the Comment paper,
but the authors of the Comment paper should be alerted.

\bigskip      
      
\subsection{Reply}

   Dear Editor,\\

Thank you for the correspondence. And we also thank the referees for their
efforts to give us the review report. But we cannot fully agree with their comments on our reply manuscript.

Now let us clarify the main point of our reply. The logic in our reply
manuscript is very simple: Firstly, in our original paper, it was shown that
the iterative method adopted in our paper to derive the generalized effective
Hamiltonian is equivalent to the Dyson series; Secondly, the truncated sum
of the Dyson series, although it is sometimes non-unitary, is the basis of the
time-dependent perturbation theory and has been extensively and successfully
employed in many fields such as quantum mechanics and quantum theory of
fields. That is to say, the non-unitary character of the truncated Dyson series
does not mean that the Dyson series is useless. Thirdly, similar to the truncated sum of the Dyson series, we think, although it might sometimes result
in non-Hermitian effective Hamiltonian, our effective Hamiltonian method can
be accepted as an approximation method as long as reasonable results can be
obtained. We believe no approximation method is completely perfect and we
think the non-Hermitian feature is not fatal as an approximation method in
particular in the small detuning case (In fact, in the actual applications the
terms with large detunings are generally ignored according to the rotating wave
approximation) and in the resonant case, the resultant effective Hamiltonian
turns out to be exactly Hermitian. However, the main point of the comment
paper is basically against the above logic. In the following we will talk about
this in response to the referees' unfair comments.

\textbf{In the Second Report of the First Referee}\\
\textbf{1. The Referee:} I'm afraid that I still do not see any new physics or arguments.\\
\textbf{Our response:} we think, as a reply to the comment, it is sufficient to point out
how our logic described above was ignored and misinterpreted by the comment
paper. The authors of the the comment actually exaggerated the influence
of non-unitary and non-Hermitian. As an approximation method, the non-Hermitian feature in our method which is just corresponding to the non-unitary
character in the truncated sum of the Dyson series is not fatal. Isn't this new
physics or arguments?\\
\textbf{2 The Referee:} The main point now seems to be that the method is equivalent
to the truncated Dyson series method, but this is already mentioned in the
comment paper.\\
\textbf{Our response:} We seriously doubt whether the referee have carefully read
both the comment and our reply manuscript. The fact is that the authors
of the comment almost ignored the equivalence between our method and the
truncated Dyson series method, the following lists from the comment paper can
tell us the truth:

(1) In the abstract:\textit{ This however, is not a very rigorous way of deriving the
effective third-order expansion for an interaction Hamiltonian with harmonic
time-dependence. In fact, here we show that the third-order Hamiltonian obtained in [1] is not Hermitian for general situations when we consider time dependence. Its non-Hermitian nature arises from the foundation of the theory
itself.}

We cannot agree with such a point, and we think our derivation is rigorous
and the non-Hermitian character can be accepted as an approximation method.

(2) In order to obtain a contradictory equation (8), the authors of the comment paper supposed the unitary time evolution described in Eqs. (6) and (7).
Apparently, they ignored the truncated Dyson series is actually non-unitary in
the case discussed in the comment paper.

(3) In the end of page 2 to page 3: \textit{However, this approach raises serious
questions because truncating the Dyson series to obtain the desired order of
effective interaction results in non unitarity [26].}

We do not agree with this point, as mentioned above, we do not think
the non-Hermitian feature in our method and the non-unitary character in the
truncated sum of the Dyson series are so serious, at least the latter was proven
to be a successful approximation method.\\
\textbf{3 The referee:} Just one comment. I do not quite follow the argument below
Eq. (6). Doesn't the condition hold if the Hamiltonian at two different time
commute?\\
\textbf{Our response:} It is very obvious that the Hamiltonian does not commute
at two different time when taking the detailed Hamiltonian of Eq. (1) in the
comment. On the other hand, the second referee agreed our argument at the
end of the review report.

So we think it is extremely unfair to give the judge failing to see the merit
in publishing the manuscript. we agree with the referee for the remark: but
perhaps a new set of eyes are needed in this case.

\textbf{In the Report of the Second Referee}\\
\textbf{1 The Referee:} Non-unitary evolution due to non-Hermitian effective Hamiltonians is an issue and this is why the Comment paper tries to fix the problem.\\
\textbf{Our response:} As mentioned above, our generalized effective Hamiltonian is
equivalent to the truncated Dyson series, thus the non-unitary time evolution
is just equivalent to the non-Hermitian effective Hamiltonian. In fact, they are
the same thing and there is no causal relationship. We admit that the authors
of the comment provided a method which solved non-Hermitian problems, but
this is also an approximation method. In our opinion, as approximation methods all have shortcomings. The non-Hermitian feature in our method and the
non-unitary character in the truncated sum of the Dyson series are not fatal
especially when the detunings are small. About such a point, we have clearly
clarified above.\\
\textbf{2 The Referee:} While the authors of the Reply claim that the non-Hermitian
nature of their effective Hamiltonians is ``no problem", they fail to provide strong
reasons to justify their view.\\
\textbf{Our response:} In our reply manuscript we never claimed the non-Hermitian
nature of the effective Hamiltonians is ``no problem", what we stated is ``one
can reasonably recognize the truncated sum of the front items in Eq. (4) as
an approximation of the effective Hamiltonian." The reason was also mentioned
in the first paragraph in page 2 ``As shown in [2], the effective Hamiltonian
expressed in Eq. (4) is equivalent to the Dyson series [6]. So the iterative
method adopted to derive the Eq. (4) is no problem and its correctness can be
guaranteed. As one knows, the Dyson series is extensively employed in many
fields such as quantum mechanics and quantum theory of fields, the truncated
sum of the Dyson series is also a basis for time-dependent perturbation theory."\\
\textbf{3 The Referee:} Specifically, when time evolution of quantum states is a topic
of interest (for example in quantum optics), an accurate description of the states
must require unitary evolution. I think that the argument based on truncated
Dyson series in the Reply does not really address the issue. Since the Reply does
not bring in something sufficiently meaningful and worthwhile to the discussion,
I do not recommend its publication.\\
\textbf{Our response:} It is sure that the accurate description of the states requires
unitary evolution, but neither our method nor the method in the comment is accurate, both methods are actually approximation method. In such approximate
cases the time evolution does not require to be unitary evolution. Otherwise,
the truncated Dyson series method would have been eliminated long ago. On
the contrary, it has been a successful method in many quantum fields, mentioned in our reply manuscript. On the other hand, our effective Hamiltonian
method was proven to be equivalent to the truncated Dyson series method, so
our method can be accepted as an approximation method even it may result in
non-Hermitian effective Hamiltonians. In the following we would like explain
the validity of our method by resorting the example in the comment paper. By
using our effective Hamiltonian method, we can derive the third-order effective
Hamiltonian
\begin{equation*}
\begin{split}
\hat{H}^{(3)}_W & = \lambda g_1g_2 \left\lbrace   \left( 1-\dfrac{\delta \hat{a}^{\dagger} \hat{a}}{\Delta_1 + \Delta_2 - \delta } \right)  \dfrac{\hat{\sigma}_{eg}\hat{a}e^{i\delta t}}{\Delta_1 (\Delta_1 + \Delta_2)}     \right. \\
& \left.   + \dfrac{\hat{a}^{\dagger} \hat{\sigma}_{ge}}{\Delta_1 + \Delta_2 - \delta }  \left[  \dfrac{1}{\Delta_1 - \delta }  - \dfrac{\delta \hat{a}^{\dagger} \hat{a}}{ \Delta_2 \left(  \Delta_1 + \Delta_2  \right) } \right] e^{-i\delta t} \right\rbrace  .
\end{split}
\end{equation*} 
where $  \delta = \Delta_1 + \Delta_2 - \omega = \omega_e - \omega $. Because  $ \delta \ll \Delta_1 $,  $ \Delta_2 $, $ H^{(3)}_W $
 can be approximately simplified as
\begin{equation*}
\begin{split}
\hat{H}^{(3)}_W & \approx \dfrac{\lambda g_1g_2 }{ \Delta_1 \left(  \Delta_1 + \Delta_2  \right) } \left[   \left( 1-\dfrac{\delta \hat{a}^{\dagger} \hat{a}}{\Delta_1 + \Delta_2 } \right) \hat{\sigma}_{eg}\hat{a}e^{i\delta t}    \right. \\
& \left.   + \hat{a}^{\dagger} \hat{\sigma}_{ge}  \left(  1- \dfrac{\Delta_1 }{\Delta_2 } \dfrac{\delta \hat{a}^{\dagger} \hat{a}}{   \Delta_1 + \Delta_2} \right) e^{-i\delta t} \right]  .
\end{split}
\end{equation*} 
The above third-order effective Hamiltonian is not Hermitian because of the presence of $ \Delta_1 / \Delta_2  $ in the term of the second line. On the other hand, considering 
$ \delta \ll \Delta_1 $,  $ \Delta_2 $, $ \dfrac{ \delta }{\Delta_1+\Delta_2} \ll 1$. In the case that $ \Delta_1 $ and $\Delta_2 $ are comparable and the photon number $ \langle  \hat{a}^{\dagger} \hat{a} \rangle $ is not very large, the terms $  \dfrac{\delta \hat{a}^{\dagger} \hat{a}}{ \Delta_1 + \Delta_2}  $
and $   \dfrac{\Delta_1 }{\Delta_2 } \dfrac{\delta \hat{a}^{\dagger} \hat{a}}{   \Delta_1 + \Delta_2} $
 have the little influence on the whole third-order effective Hamiltonian $ \hat{H}^{(3)}_W $.
 Moreover, as $  \delta$ deceases their influences reduce and when $ \delta = 0 $ $ \hat{H}^{(3)}_W $ becomes
\begin{equation*}
\hat{H}^{(3)}_W = \dfrac{\lambda g_1g_2 }{ \Delta_1 \left(  \Delta_1 + \Delta_2  \right) }    \left(    \hat{\sigma}_{eg}\hat{a}    + \hat{a}^{\dagger} \hat{\sigma}_{ge} \right) .
\end{equation*}   

Obviously, it is Hermitian and photon number independent in this case. Here
we find another mistake appearing in the comment paper, that is, the Eq. (27)
in the comment paper is different from the above Hamiltonian.

Based on the above response to the referees, we strongly suggest to accept
our reply for publication.\\

Yours sincerely,

The authors

\newpage

\section{Decision and appeal}

\subsection{Decision}
 
 Dear Dr. Feng,\\

We consulted one of our Editorial Board members regarding your appeal of the 
decision on this manuscript.  Please find below the report we received in 
response, which upholds the rejection. In accordance with our standard practice, 
this concludes the scientific review of your manuscript.\\

Yours sincerely,

Dr. *** ***

Managing Editor

Physical Review A

Email: pra@aps.org

https://journals.aps.org/pra/

Follow us on Twitter @PhysRevA\\
--------------------------------------------------------------------------
Report of the Editorial Board Member -- AHK1043/Shao
--------------------------------------------------------------------------

Having reviewed the correspondence between the authors of the reply to the
comment and the referees, I agree with the editor’s decision to reject the
publication of the reply. 

I will point out that my support of the rejection to the reply is not meant to
imply that the authors of the reply are wrong in their arguments or conversely
the that the contents of the Comment on the original paper is somehow wrong
(this I suppose is for the scientific community to decide). 
My recommendation is based solely on the fact that the reply does not seem to
add any new physics (or for that matter a new point of view) to the proceedings.\\

Prof. *** ***

Editorial Board Member

Physical Review A

\bigskip      
      
\subsection*{Appeal}    
  
Dear Committee,\\

We thank the Editorial Board member for giving an answer to the appeal review. It is very clear that the iterative method used to obtain the third-order effective Hamiltonian in our and James' paper is correct, and there are some new physics in our reply, in which the truncated Dyson series can be utilized in the time-dependent perturbation theory.

Moreover, the comment with being second revised still has several mistakes and minor problems.  It needs to be pointed out that if the mistakes in the comment paper can not be corrected, it will not only be unfair to our paper, but also seriously mislead the readers. This is a lack of strict requirements and fairness for PRA, it will disappoint the authors and readers of the Journal. 

By the way,  a reply should answer the question or give the explanation to that the comment raises or has, so are new physics and arguments needed? And the referees' reports are not to the point (makes wrong conclusion about our work) and do not give sufficient reasons to reject our reply.  We think that the reply has made explanation and correction to the comment, and suggest  that the reply can be accepted for publication. \\

Yours sincerely,

The authors


\begin{thebibliography}{9}
\bibitem{comment} W. Rosado, and I. Arraut, Comment on ``Generalized James' Effective Hamiltonian Method'', Phys. Rev. A \textbf{108}, 066201 (2023).

\bibitem{Shao} W. Shao, C. Wu, and X.-L. Feng, Generalized James' effective Hamiltonian method, Phys. Rev. A \textbf{95}, 032124 (2017).

\bibitem{James-2000} D. F. V. James, Quantum computation with hot and cold ions: An assessment of proposed schemes, Fortschr. Phys. \textbf{48}, 823 (2000).

\bibitem{James-2007} D. F. V. James and J. Jerke, Effective Hamiltonian theory and its applications in quantum information, Can. J. Phys. \textbf{85}, 625 (2007).

\bibitem{James-2010} O. Gamel and D. F. V. James, Time-averaged quantum dynamics and the validity of the effective Hamiltonian model, Phys. Rev. A \textbf{82}, 052106 (2010).

\bibitem{Dyson} F. J. Dyson, The radiation theories of Tomonaga, Schwinger, and Feynman, Phys. Rev. \textbf{75}, 486 (1949).

\bibitem{ME} S. Blanes, F. Casas, J. A. Oteo, J. Ros, The Magnus expansion and some of its applications, Phys. Rep. \textbf{470,} 151 (2009).
\end{thebibliography}
\end{document}